\documentstyle[twoside]{article}
\catcode`\@=11
\long\def\@makefntext#1{
\protect\noindent \hbox to 3.2pt {\hskip-.9pt
$^{{\eightrm\@thefnmark}}$\hfil}#1\hfill}		

\def\@makefnmark{\hbox to 0pt{$^{\@thefnmark}$\hss}}	

\def\ps@myheadings{\let\@mkboth\@gobbletwo
\def\@oddhead{\hbox{}
\rightmark\hfil\eightrm\thepage}
\def\@oddfoot{}\def\@evenhead{\eightrm\thepage\hfil
\leftmark\hbox{}}\def\@evenfoot{}
\def\sectionmark##1{}\def\subsectionmark##1{}}
\oddsidemargin=\evensidemargin
\newcounter{sectionc}\newcounter{subsectionc}\newcounter{subsubsectionc}
\renewcommand{\section}[1] {\vspace{12pt}\addtocounter{sectionc}{1}
\setcounter{subsectionc}{0}\setcounter{subsubsectionc}{0}\noindent
	{\tenbf\thesectionc. #1}\par\vspace{5pt}}
\renewcommand{\subsection}[1] {\vspace{12pt}\addtocounter{subsectionc}{1}
	\setcounter{subsubsectionc}{0}\noindent
	{\bf\thesectionc.\thesubsectionc. {\kern1pt \bfit #1}}\par\vspace{5pt}}
\renewcommand{\subsubsection}[1] {\vspace{12pt}\addtocounter{subsubsectionc}{1}
	\noindent{\tenrm\thesectionc.\thesubsectionc.\thesubsubsectionc.
	{\kern1pt \tenit #1}}\par\vspace{5pt}}
\newcommand{\nonumsection}[1] {\vspace{12pt}\noindent{\tenbf #1}
	\par\vspace{5pt}}
\newcommand{\textlineskip}{\baselineskip=13pt}
\newcommand{\smalllineskip}{\baselineskip=10pt}
\def\eightcirc{
\begin{picture}(0,0)
\put(4.4,1.8){\circle{6.5}}
\end{picture}}
\def\eightcopyright{\eightcirc\kern2.7pt\hbox{\eightrm c}}
\newcommand{\copyrightheading}[1]
	{\vspace*{-2.5cm}\smalllineskip{\flushleft
	{\footnotesize International Journal of Modern Physics A, #1}\\
	{\footnotesize $\eightcopyright$\, World Scientific Publishing
	 Company}\\
	 }}
\newcommand{\pub}[1]{{\begin{center}\footnotesize\smalllineskip
	Received #1\\
	\end{center}
	}}

\def\abstracts#1#2#3{{
	\centering{\begin{minipage}{4.5in}\baselineskip=10pt\footnotesize
	\parindent=0pt #1\par
	\parindent=15pt #2\par
	\parindent=15pt #3
	\end{minipage}}\par}}

\newcommand{\bibit}{\nineit}
\newcommand{\bibbf}{\ninebf}
\renewenvironment{thebibliography}[1]
	{\frenchspacing
	 \ninerm\baselineskip=11pt
	 \begin{list}{\arabic{enumi}.}
	{\usecounter{enumi}\setlength{\parsep}{0pt}
	 \setlength{\leftmargin 12.7pt}{\rightmargin 0pt} 
	 \setlength{\itemsep}{0pt} \settowidth
	{\labelwidth}{#1.}\sloppy}}{\end{list}}
\newcounter{itemlistc}
\newcounter{romanlistc}
\newcounter{alphlistc}
\newcounter{arabiclistc}
\newenvironment{itemlist}
    	{\setcounter{itemlistc}{0}
	 \begin{list}{$\bullet$}
	{\usecounter{itemlistc}
	 \setlength{\parsep}{0pt}
	 \setlength{\itemsep}{0pt}}}{\end{list}}
\newenvironment{romanlist}
	{\setcounter{romanlistc}{0}
	 \begin{list}{$($\roman{romanlistc}$)$}
	{\usecounter{romanlistc}
	 \setlength{\parsep}{0pt}
	 \setlength{\itemsep}{0pt}}}{\end{list}}

\def\@citex[#1]#2{\if@filesw\immediate\write\@auxout
	{\string\citation{#2}}\fi
\def\@citea{}\@cite{\@for\@citeb:=#2\do
	{\@citea\def\@citea{,}\@ifundefined
	{b@\@citeb}{{\bf ?}\@warning
	{Citation `\@citeb' on page \thepage \space undefined}}
	{\csname b@\@citeb\endcsname}}}{#1}}
\newif\if@cghi
\def\cite{\@cghitrue\@ifnextchar [{\@tempswatrue
	\@citex}{\@tempswafalse\@citex[]}}
\def\citelow{\@cghifalse\@ifnextchar [{\@tempswatrue
	\@citex}{\@tempswafalse\@citex[]}}
\def\@cite#1#2{{$\null^{#1}$\if@tempswa\typeout
	{IJCGA warning: optional citation argument
	ignored: `#2'} \fi}}
\newcommand{\citeup}{\cite}
\def\pmb#1{\setbox0=\hbox{#1}
	\kern-.025em\copy0\kern-\wd0
	\kern.05em\copy0\kern-\wd0
	\kern-.025em\raise.0433em\box0}

\def\fnt#1#2{\footnotetext{\kern-.3em
	{$^{\mbox{\scriptsize #1}}$}{#2}}}
\def\fpage#1{\begingroup
\voffset=.3in
\thispagestyle{empty}\begin{table}[b]\centerline{\footnotesize #1}
	\end{table}\endgroup}
\def\runninghead#1#2{\pagestyle{myheadings}
\markboth{{\protect\footnotesize\it{\quad #1}}\hfill}
{\hfill{\protect\footnotesize\it{#2\quad}}}}
\font\tenrm=cmr10
\font\tenit=cmti10
\font\tenbf=cmbx10
\font\bfit=cmbxti10 at 10pt
\font\ninerm=cmr9
\font\nineit=cmti9
\font\ninebf=cmbx9
\font\eightrm=cmr8

\begin{document}

\runninghead{The Exact Cosmological Solution to the Dynamical Equations
for the Bianchi IX Model}
{The Exact Cosmological Solution to the Dynamical Equations
for the Bianchi IX Model}

\normalsize\textlineskip
\thispagestyle{empty}
\setcounter{page}{1}

\copyrightheading{}			

\vspace*{0.88truein}

\fpage{1}
\centerline{\bf THE EXACT COSMOLOGICAL SOLUTION}
\vspace*{0.035truein}
\centerline{\bf TO THE DYNAMICAL EQUATIONS}
\vspace*{0.035truein}
\centerline{\bf FOR THE BIANCHI-IX MODEL}
\vspace*{0.37truein}
\centerline{\footnotesize V.A. SAVCHENKO\footnote{e-mail:
savchenko@phys.rnd.runnet.ru}, T.P. SHESTAKOVA\footnote{e-mail:
stp@phys.rnd.runnet.ru}, G.M. VERESHKOV}
\vspace*{0.015truein}
\centerline{\footnotesize\it Department of theoretical physics, Rostov State
University}
\baselineskip=10pt
\centerline{\footnotesize\it Sorge str. 5, Rostov-on-Don 344090, Russia}
\vspace*{0.225truein}
\pub{5 April 1999}

\vspace*{0.21truein}
\abstracts{ Quantum geometrodynamics in extended phase space describes
phenomenologically the integrated system ``a physical object + observation
means (a gravitational vacuum condensate)". The central place in this version
of
QGD belongs to the Schr\"odinger equation for a wave function of the
Universe. An exact solution to the ``conditionally-classical" set of
equations in extended phase space for the Bianchi-IX model and the
appropriate solution to the Schr\"odinger equation are considered. The
physical adequacy of the obtained solutions to existing concepts about
possible cosmological scenarios is demonstrated. The gravitational vacuum
condensate is shown to be a cosmological evolution factor.}{}{}

\vspace*{1pt}\textlineskip

\section{Introduction}

\label{Intro}

\vspace*{-0.5pt}
\noindent
In the previous paper\citeup{SSV} we studied quantum geometrodynamics (QGD)
of the Bian\-chi-IX model in the framework of extended phase space (EPS)
approach elaborated by Batalin, Fradkin and Vilkovisky
(BFV)\citeup{BFV1}$^-$\citeup{BFV5}.
It is generally accepted that in the BFV approach one singles out a
BRST-invariant sector which is supposed to coincide with a gauge-invariant
one. This scheme was realized in the works\citeup{T}$^-$\citeup{HH}; it has a manifest
physical interpretation in the case of S-matrix theory. However, in
quantum cosmology appropriate mathematical operations are just
formal and the question arises, whether these operations are mathematically and
physically justified. This question has been explored in our
paper\citeup{SSV}. Our main result consisted in the demonstration that in
a closed universe without asymptotical states BRST-invariance is not
equivalent to gauge invariance. A mathematical indication to the
nonequivalence of gauge and BRST invariance is given by the well-known
parametrization noninvariance of the Wheeler -- DeWitt (WDW) equation.
As we have shown in the framework of the BFV approach, the latter is
an ill-hidden gauge noninvariance. For this reason singling out a
BRST-invariant sector is not motivated in QGD and is not used in our
approach. This circumstance distinguished our version of QGD from other
works on quantum geometrodynamics. A physical ground for our approach
consists in the impossibility to remove an observer from a closed universe,
and, as a consequence, the necessity to take into account his affecting
physical processes. The lack of gauge symmetry, or, more precisely,
breaking down this symmetry when considering observation means (OM)
as a part of an integrated system in our version of QGD is described by
eigenvalues of gravitational super-Hamiltonian; we refer to the latters as
energy levels of gravitation vacuum condensate (GVC).

It is worth emphasizing that in a gauge-invariant approach to QGD it is
assumed that the information about the reduction of a wave packet in the
process of evolution of the Universe is contained in boundary
conditions\citeup{Halliwell}$^-$\citeup{HL}. In contrast to the
conventional approach, our version of QGD admits the Copenhagen
interpretation and aims at describing the integrated system ``a physical
object (gravitational field) + observation means (gravitational vacuum
condensate)''.

Approaches of quantum cosmology are traditionally tested and developed
for the Bianchi-IX model and its particular case -- the isotropic
model\citeup{Marolf}$^-$\citeup{Kodama}. As it is known, the Bianchi-IX
model can be represented as an isotropic space where two transversal
nonlinear gravitational waves are excited. So, in papers based on the
conventional Wheeler -- DeWitt QGD (or in those where the BRST-invariant
sector in EPS is singled out) observables include 3-space volume and
amplitudes of gravitational waves. The convential approach is developing
applying new consepts such as supergavity and
superstrings\citeup{Graham}$^-$\citeup{Lidsey}.
However, the interpretation of the Wheeler -- De Witt theory is not
changing: the information about the past of the Universe as well as
its future is contained in boundary conditions. In our modification of QGD
a  new feature appears -- the gravitational vacuum condensate with its
degrees of freedom. Wave functions of the integrated system are represented
by normalized wave packets; we suppose that their structure is determined
while the Universe creating from ``Nothing''. The influence of observation
means on the evolution of the integrated system in the proposed version of
QGD is described by transformation of the form of a wave packet.

The wave functions satisfy the dynamical Schr\"odinger equation (SE).
For the Bianchi-IX model the normalized in the EPS general solution (GS)
to the gauge-noninvariant SE reads
\begin{equation}
\label{time-depend.WF}
\Psi(Q^a,Q^0,\theta,\bar{\theta};t)=\int\Psi_k(Q^a)\exp(-iE_kt)
(\bar{\theta}+i\theta)\delta(Q^0-f(Q^a)-k)\,dE_kdk,
\end{equation}
where $\Psi_k(Q^a)$ is a solution to the stationary equation
\begin{equation}
\label{station.phys.SE}
H^0_k\Psi_k(Q^a)=E_k\Psi_k(Q^a),
\end{equation}
$\theta, \bar\theta$ are the ghosts, $Q^0$ and $Q^a$ are the gauge and
physical variables respectively, specification of which see below in Sec.~2.

The norm integral for the WF (\ref{time-depend.WF}) over the full set of
variables $Q,\theta,\bar\theta$ results in
$$
\int\Psi^*_k(Q^a,t)\Psi_k(Q^a,t)M_k(Q^a)\,dk\prod_adQ^a,
$$
so the GS to the SE, under the condition the $\Psi_k(Q^a,t)$ to be a
sufficiently narrow packet over $k$, is normalizable with respect to the
gauge variable, as well as to the ghosts and the physical variables.

The theory does not control the WF dependence on the parameter $k$ which in
the classical dynamics determines an initial clock setting, and this
additional degree of freedom has to be referred to an observer.

Investigation of the particular BRST-invariant solution has shown that it
really satisfies the WDW equation but cannot be normalized. On the other
hand, as we have already mentioned above, the well known parametrization
noninvariance of the WDW equations is shown to be the ill-hidden {\it gauge}
noninvariance. The latter means that the WDW theory does not achieve its
object to give a gauge-invariant description of the Universe.

By this reason and because of the loss of probability interpretation we do not
see any ground to be guided by the WDW theory, and explore the
possibilities of the gauge-noninvariant QGD version.

The factored part of the GS (\ref{time-depend.WF}) -- $\delta$-function and
ghosts -- represents the OM described by the gauge-fixing term in the
Lagrangian and by the appropriate ``energy-momentum tensor" (quasi-EMT).
The latter corresponds to the continuous medium that fills the whole
space and can be called ``gravitational vacuum condensate"; it is
quantitatively fixed by an eigenvalue $E_k$ of the Hamiltonian $H^0_k$. The
structure of the GS shows a GVC to be an important factor of the global
evolution. The purpose of this paper is to investigate this effect on the base
of the exact solution to the SE for the simplified model with the frozen
degree of freedom $Q^3$, in a simple gauge.

In Sec.\,\ref{Model}. all the necessary notations and the general equations
for the
Bianchi-IX model are given; in Sec.\,\ref{Class}. the exact
conditionally-classical
solution for the simple case is considered. We shall show that GVC can play
a decisive role in forming non-trivial cosmological scenarios. In 
Sec.\,\ref{Quant}. the 
exact solution to the SE is considered, we analyse the formation and 
properties of the wave packet representing the evolving universe. In 
Sec.\,\ref{Semiclass}. the transition to the semiclassical solution is 
considered and it is shown how the quantum cosmological effects become 
negligible in the semiclassical region; here various cosmological scenarios depending on the GVC and the 
homogeneous scalar field are considered. In Sec.\,\ref{time}. the concept of 
time in the gauge-noninvariant QGD is discussed. 

\section{The General Model Equations}
\label{Model}
\noindent
The Bianchi-IX 4-interval is given by
\begin{equation}
\label{ds}
ds^2=N^2(t)\,dt^2-\eta_{ab}(t)e^a_ie^b_k\,dx^idx^k;
\end{equation}
\begin{equation}
\label{eta_ab}
\eta_{ab}(t)={\,\rm diag}\left(a^2(t),b^2(t),c^2(t)\right),
\end{equation}
$$
\begin{array}{l}
e^1_i=(\sin x^3,-\cos x^3\sin x^1,0),\\
e^2_i=\rule{0pt}{14pt}(\cos x^3,\sin x^3\sin x^1,0),\\
e^3_i=\rule{0pt}{14pt}(0,\cos x^1,1).
\end{array}
$$
$$
a=\frac12 \exp\left[\frac12\left(Q^1+Q^2+\sqrt{3}\,Q^3\right)\right];\; 
b=\frac12 
\exp\left[\frac12\left(Q^1+Q^2-\sqrt{3}\,Q^3\right)\right];\;
$$
$$
c=\frac12 \exp\left(\frac12Q^1-Q^3\right).
$$
The model is assumed to include an arbitrary number $K$ of real 
homogeneous scalar fields $Q^4,\ldots, Q^{K+3}$ with some potential $U_s$. The 
gauge coordinate $Q^0$ is defined by an arbitrary parametrization function
\begin{equation}
\label{zeta}
\zeta(Q^0,Q^1,\ldots,Q^{K+3})=\ln\left[\frac1N\exp\left(\frac32Q^1\right)
\right],
\end{equation}
and gauges are supposed to be not depending on time,
\begin{equation}
\label{gauge_class}
Q^0=f(Q^1,\ldots)+k, k={\,\rm const}.
\end{equation}

The SE reads
\begin{equation}
 \label{SE1}
i\frac{\partial\Psi(Q^0,Q^1,\ldots,\theta,\bar{\theta};t)}{\partial 
t}=H\Psi(Q^0,Q^1,\ldots,\theta,\bar{\theta};t), 
\end{equation}
where $\Psi(Q^\alpha,\theta,\bar\theta;t)$ is a universe WF 
$(\alpha=0,a;a=1,\ldots,K+3), \theta,\bar\theta$, are the ghosts,
\begin{equation}
\label{H_full}
H=- i\zeta,_0\frac{\partial}{\partial\theta} 
\frac{\partial}{\partial\bar{\theta}}- 
\frac1{2M}\frac{\partial}{\partial Q^{\alpha}}G^{\alpha\beta}
\frac{\partial}{\partial Q^{\beta}}+{\,\rm e}^{-\zeta}(U-V), 
\end{equation}
$\zeta,_0=\partial\zeta(Q^\alpha)/\partial Q^0$, probability measure
\begin{equation}
\label{zeta,M-coupl}
M={\,\rm const}\cdot \zeta,_0\exp\left(\frac{K+3}2\zeta\right),
\end{equation}
\begin{equation}
\label{U,Ug,Us}
U(Q)={\,\rm e}^{2Q^1}U_g(Q^2,Q^3)+{\,\rm e}^{3Q^1}U_s(Q^4,\ldots), 
\end{equation}
\begin{eqnarray*}
U_g(Q^2,Q^3)&=&\frac23\left\{\exp\left[2\left(Q^2+\sqrt{3}\,Q^3 
\right)\right]+\exp\left[2\left(Q^2-\sqrt{3}\,Q^3\right)\right]+\exp(-
4\chi)\right.\\&-&2\exp\left[-\left(Q^2+\sqrt{3}\,
Q^3\right)\right]-\left.2\exp\left(-Q^2+\sqrt{3}\,Q^3\right)-
2\exp(2Q^2)\right\}
\end{eqnarray*}
\begin{eqnarray}
\label{V}
V=&-&\frac3{12}\frac{(\zeta,_0)^a(\zeta,_0)_a}{\zeta,_0\!\!\!^2}+
\frac{(\zeta,_0)^a_a}{3\zeta,_0}+\frac{K+1}{6\zeta,_0}\zeta_a 
(\zeta,_0)^a\nonumber\\&+&\frac1{24}(K^2+3K+14)\zeta_a\zeta^a 
+\frac{K+2}6\zeta_a^a, 
\end{eqnarray}
$\zeta_a=\partial \zeta/\partial Q^a+f,_a\partial \zeta/\partial Q^0$,
$$
G^{\alpha\beta}=M{\,\rm e}^{-\zeta}\left(
\begin{array}{cc}
f,_a\!f^{,a}&f^{,a}\\
f^{,a}&\gamma^{ab}
\end{array}
\right).
$$

To make the analysis more visual, we begin with considering the {\it 
conditionally}-classical solution to the Einstein equations supplemented with 
the quasi-EMT of the ghosts and the GVC. In common case this set of 
equations, yielded by the ``classical version" of the Hamiltonian 
(\ref{H_full}), reads
\begin{eqnarray}
\label{Q-eqn}
\left({\,\rm e}^\zeta\dot{Q}_a\right)^.&+&{\,\rm e}^{-\zeta}U,_a-
\dot{\lambda}f,_a-\frac12 
\zeta,_a{\,\rm e}^\zeta \dot{Q}^b\dot{Q}_b-
\zeta,_a{\,\rm e}^{\zeta}U+\frac{i\zeta,_0,_a}
{\zeta,_0^2}\dot{\bar{\theta}}\dot{\theta}=0,\\
\label{0-eqn}
\frac12\zeta,_0{\,\rm e}^\zeta\dot{Q}^a\dot{Q}_a&+&\zeta,_0{\,\rm e}^{-
\zeta}U-
\frac{i\zeta,_0,_0}{\zeta,_0^2}\dot{\bar{\theta}}\dot{\theta}-
\dot{\lambda}=0, \\
\label{pi-eqn}
\dot Q^0-f,_a\dot Q^a&=&0, \\
\label{bartheta-eqn}
\left(\zeta,_0^{-1}\dot{\theta}\right)^.&=&0,\\ 
\label{theta-eqn}
\left(\zeta,_0^{-1}\dot{\bar{\theta}}\right)^.&=&0, 
\end{eqnarray}
where $\lambda=\pi+\dot{\bar{\theta}}\theta,\;\pi$ is the Lagrange multiple 
fixing the gauge (\ref{pi-eqn}), coupled to the integral of motion describing 
the GVC: 
$$
(\zeta,_0)_k^{-1}\dot{\lambda}=E_k. 
$$

\section{The Conditionally-Classical Exact Solution}
\label{Class}
\noindent
Taking the parametrization and the gauge 
\begin{equation}
\label{simplegauge}
\zeta=Q^0=k 
\end{equation}
($k={\,\rm const}$), one can obtain an exact particular solution to 
Eqs.~(\ref{Q-eqn}) -- (\ref{theta-eqn}) with $Q^3=0$. The existence of this 
solution gives the formal opportunity to consider the model without this 
degree of freedom.
	
Under the condition (\ref{simplegauge}) the ghost variables vanish from 
Eqs.~(\ref{Q-eqn}) -- (\ref{pi-eqn}), and the latter form a closed set 
concerning the physical variables; the state equation of the GVC becomes 
extremely hard,
\begin{equation}
\label{GVC_state_eqn}
p=\varepsilon=-\frac{\dot{\lambda}}{2\pi^2}\exp(k-3Q^1), 
\end{equation}
\begin{equation}
\label{lamb,E}
\dot{\lambda}=E. 
\end{equation}
	   
Note that conditionality of the classical approach shows here in the ghost 
presence in the integral of motion (\ref{lamb,E}) 
$$\dot{\lambda}=\dot{\pi}-\dot{\bar{\theta}}\dot{\theta},$$
i.e. the forms made of the Grassmannian variables appear as parameters of 
the theory.

Let us turn to the case of a single massless linear scalar field $\phi=Q^4$ and 
put 
$$
U_s(\phi)=0, 
$$
in (\ref{U,Ug,Us}). Now we have the simple equation for $\phi$
$$\ddot{\phi}=0,$$
thus,
\begin{equation}
\label{dotphi,Cs}
\dot{\phi}=C_s={\,\rm const}.
\end{equation}
The scalar field behaves as a medium with positive energy density and with 
an extremely hard equation of state
$$
p_{(scal)}=\varepsilon_{(scal)}\propto\exp(-
3Q^1)\dot{\phi}^2=C^2_s\exp(-3Q^1)
$$
like that of the GVC (\ref{GVC_state_eqn}).

As one can see, in the present model the Universe is filled with the 
two-compo\-nent medium described by the parameters $E$
and $C_s$. Below we will show  that relation between the two parameters 
essentially affects cosmological evolution at the quantum stage of the 
Universe existence as well as at the semiclassical one. Here is the 
difference between our consideration and the usual investigation of the 
Bianchi-IX model in general relativity.

The equations for $Q^1,Q^2$ take the form: 
$$
\ddot Q^1-\frac43\left[\exp(2Q^1-4Q^2)-4\exp(2Q^1-Q^2)\right]=0,
$$
$$
\ddot{Q^2}-\frac43[2\exp(2Q^1-4Q^2)-2\exp(2Q^1-Q^2)]=0.
$$
Integration is simplified with the substitution
\begin{equation}
\label{z1,z2}
z_1=2Q^1-4Q^2,\;\;z_2=2Q^1-Q^2;
\end{equation}
after replacing
$$
t\to {\,\rm e}^{-k}t 
$$
the solution is available in the form
\begin{equation}
\label{solutions}
\exp\left(\frac{z_2}2\right)=\displaystyle\frac{\alpha}{\cosh[2\alpha(t-
t_0)]},\;\;
\exp\left(\frac{z_1}2\right)=\displaystyle\frac{\vphantom{\hat{A}}\beta}
{\cosh[2\beta (t-t_1)]},
\end{equation}
where $\alpha,\beta,t_0,t_1$ are the integration constants. Without loss of 
generality, by shifting zero time one can put $t_1=0$. For the metric 
(\ref{ds}) -- (\ref{eta_ab}) one finds:
$$
a^2=b^2=\frac14\exp\left(z_2-\frac12z_1\right)=\frac{a^2\cosh(2\beta t)}
{4\beta\cosh^2[2\alpha(t-t_0)]};
$$
$$
c^2=\frac14\exp\left(\frac12z_1\right)=\frac{\beta}{4\cosh(2\beta t)}.
$$
From the constraint equation (\ref{0-eqn}) with $Q^3=0$, and 
(\ref{dotphi,Cs}) it follows:

$$
\frac1{24}\left(\dot z_1^2-4\dot z_2^2\right)+\frac23\left[\exp(z_1)
-4\exp(z_2)\right]= E_k-\frac12C^2_s, 
$$
where $E_k={\,\rm e}^kE$; hence, in turn,
\begin{equation}
\label{alpha,beta-invar}
\alpha^2-\frac14\beta^2=-\frac38\left(E_k-\frac12C^2_s\right).
\end{equation}
So, the dynamics of the model depends qualitatively on a relation between 
$C_s$ and $E_k$. Various cosmological effects of this dependence will be 
discussed in Sec.~5., after considering the quantum version of the exact 
solution.

\section{The Exact Solution to the Schr\"odinger Equation} 
\label{Quant}
\noindent
The task of constructing WF (\ref{time-depend.WF}) is reduced to searching 
for a solution to stationary Eq.~(\ref{station.phys.SE}) for the physical part 
of the WF under the parametrization-and-gauge condition 
(\ref{simplegauge}). This equation reads
\begin{equation}
\label{SEphys}
-\frac12\frac{\partial^2\Psi_k}{\partial Q_a\partial Q^a} 
+U(Q^a)\Psi_k(Q^a)=E_k\Psi_k(Q^a),
\end{equation}
$a=(1,2,4)$. Substitution of (\ref{z1,z2}) enables to separate the 
variables in the equation, and it can be written in the following manner
$$
\left(6\hat{L}_1-\frac32\hat{L}_2+\frac12\hat{L}_3-
kE\right)\Psi_k(z_1,z_2,\phi)
=0,
$$
where
\begin{eqnarray*}
\hat{L}_1&=&-\frac{\partial^2}{\partial z_1^2}+\frac19\exp(z_1),\\
\hat{L}_2&=&-\frac{\partial^2}{\partial z_2^2}+\frac{16}9\exp(z_2),\\
\hat{L}_3&=&-\frac{\partial^2}{\partial\phi^2}.
\end{eqnarray*}

The eigenfunctions of the operators $\hat{L}_1,\hat{L}_2$ appropriate to the
positive eigenvalues $\nu^2_1/4,\nu^2_2/4$ are the modified Bessel 
functions with an imaginary index,
$$
\psi_{\nu}(z)=\frac1{\sqrt{2\pi\,\Gamma(i\nu)}}K_{i\nu}
\left[A\exp\left(\frac 
z2\right)\right];\;\;(A_1;A_2)=\left(\frac23;\frac83\right).$$

So far as
$$
\exp\left(\frac{z_1}2\right)=4c^2;\;\;\exp\left(\frac{z_2}2\right)=4ac,
$$
the quantum number $\nu_1$ determines probability distribution for the scale 
$c$, and so does the quantum number $\nu_2$ for the scale $a=b$ at a given 
$c$ value. Note, that for any $\nu$ there exists a semiclassical solution to the 
problem,
\begin{eqnarray*}
\psi_{\nu}(z)&=&\displaystyle\frac1{\sqrt2\,\Gamma(i\nu)\exp\left( 
\displaystyle\frac{\pi\nu}2\right)\left(\displaystyle\frac{\nu^2}4-
\displaystyle\frac{A^2}4\exp(z)\right)^{\frac14}}\cos\left[2\left(\sqrt{\frac 
{\nu^2}4-\frac{A^2}4\exp(z)}\right.
\right.\\&-&\left.\left.\frac{\nu}2\mathop{\rm Artanh}\sqrt{1-
\frac{A^2}{\nu^2}\exp(z)}\right)+\frac{\pi}4\right],
\end{eqnarray*}
$z<z_{\nu}$;
\begin{eqnarray*}
\psi_{\nu}(z)&=&\displaystyle\frac1{2\sqrt{2}\,\Gamma(i\nu)\exp\left(
\displaystyle\frac{\pi\nu}2\right)\left(\displaystyle\frac{A^2}4\exp(z)-
\displaystyle\frac{\nu^2}4\right)^{\frac14}}
\exp\left[-2\left(\sqrt{\frac{A^2}4\exp(z)-\frac{\nu^2}4}\right.
\right.\\&-&\left.\left.\frac{\nu}2\arctan\sqrt{\frac{A^2}{\nu^2}\exp(z)
-1}\right)\right],
\end{eqnarray*}
$z>z_{\nu}, z_{\nu}=\ln(\nu^2/A^2)$ being the classical turning-point.

The operator $\hat{L}_3$ eigenfunctions are the plane waves
$$
\psi_\varrho(\phi)=\frac1{\sqrt{2\pi}}\exp(i\varrho\phi)
$$
that is in agreement with the classical solution (\ref{dotphi,Cs}).

The GS to Eq.~(\ref{SEphys}) for a given value of the parameter $E_k$, 
describing a GVC state, is a superposition
\begin{eqnarray}
\label{GSexact}
\Psi_{E_k}(z_1,z_2,\phi)&=&\int\!\!\!\!\int\limits_{-
\infty}^{\infty}\!\!\!\!\int d\varrho d\nu_1d\nu_2c_1(\nu_1,\nu_2,\varrho) 
\psi_{\nu_1}(z_1) \psi_{\nu_2}(z_2)\psi_\varrho(\phi)\nonumber\\ 
&\times&\delta\left(\frac32\nu^2_1-\frac38\nu^2_2+\frac12\varrho^2-E_k\right).
\end{eqnarray}

However, the stationary states, the wave functions (\ref{GSexact}) 
correspond to, are not physical, being unnormalizable because of continuity 
of the $E_k$ value spectrum. A time-dependent wave packet fits a physical 
state:
\begin{equation}
\label{exactE-pack}
{\Psi_k(z_1,z_2,\phi,t)=\int\limits_{-\infty}^{\infty}dE_kc_2(E_k)}
\Psi_{E_k}(z_1,z_2,\phi)\exp\left[-iE_k(t-t_0)\right].
\end{equation}

Note, that in the expressions (\ref{GSexact}), (\ref{exactE-pack}) the 
quantity $E_k$ appears as a controlling parameter providing, via the 
$\delta$-function, correlation of the quantum numbers $\nu_1,\nu_2,\varrho$, and, 
by that, a probability distribution of the space scales at the quantum stage of 
the Universe evolution.

\section{The Semiclassical Regime and Cosmological Effects}
\label{Semiclass}
\noindent 
One can obtain the classical evolution law by computing  the operators 
$\exp(z_1/2)$, $\exp(z_2/2)$ mean values over the packet 
(\ref{exactE-pack}). 
The matrix elements will be required for that,
\clearpage
\begin{eqnarray}
\label{matr.elem}
\lefteqn{\int\limits_{-\infty}^{\infty}dz\exp\left(\frac 
z2\right)\psi^*_{\mu}(z)
\psi_{\nu}(z)}\nonumber\\&=&\left[2\pi\Gamma(-
i\mu)\Gamma(i\nu)\right]^{-1}
\int\limits_{-\infty}^{\infty}dz\exp\left(\frac z2\right)K_{-i\mu}\left[A\exp
\left(\frac z2\right)\right]K_{i\nu}\left[A\exp\left(\frac 
z2\right)\right]\nonumber\\
&=&\pi\left[4A\Gamma(-i\mu)\Gamma(i\nu)\right]^{-1}
\left\{\cosh\left[\frac{\pi}2(\mu+\nu)\right]\cosh\left[\frac{\pi}2
(\mu-\nu)\right]\right\}^{-1}.
\end{eqnarray}

To be able to describe really the classically evolving Universe, the packet 
(\ref{GSexact}) -- (\ref{exactE-pack}) should be sufficiently narrow, i.e. 
$c_1(\nu_1,\nu_2,\varrho)$ and $c_2(E_k)$ should not deviate from zero values 
beyond a small vicinity of their arguments near 
$(\bar{\nu}_1,\bar{\nu}_2,\bar\varrho)$ 
and $\bar{E}_k$. Therefore,
\begin{equation}
\label{mu,nu,omega} 
\mu+\nu\approx2\bar{\nu};\;\;\mu-\nu\approx\frac{A\omega}{2\bar{\nu}},
\end{equation}
where $\omega=A^{-1}(\mu^2-\nu^2)$ is the difference between the two 
values of the parameter $E_k$ corresponding to the quantum numbers 
$\mu$ and $\nu$. Note that the matrix element (\ref{matr.elem}) depends 
weakly on $\nu$ and is sharply decreasing when $|\mu-\nu|$ increasing. And 
so, making use of the approximations (\ref{mu,nu,omega}) one obtaines, for 
the average exponents $\exp(z/2)$,
\begin{eqnarray}
\label{aver.exp}
\overline{\exp\left(\frac z2\right)}&=&\frac14\tanh(\pi\bar{\nu})
\int\limits_{-\infty}^{\infty}d\omega\displaystyle\frac{\exp\left[-i\omega(t-
t_0)\right]}{\cosh\left(\displaystyle\frac{\pi A\omega}{4\bar{\nu}}\right)}
\nonumber\\&=&\displaystyle\frac{\bar{\nu}\tanh(\pi\bar{\nu})}
{A\cosh\left[(2A^{-1}\bar{\nu}(t-t_0)\right]}. 
\end{eqnarray}
In the classical limit $\bar{\nu}$ is large, hence 
$\tanh(\pi\bar{\nu})\approx1$, and, comparing (\ref{aver.exp}) with the 
classical expression (\ref{solutions}), one concludes that
$$
\alpha=\frac{\bar{\nu}_2}{A_2}=\frac38\bar{\nu}_2;\;\;
\beta=\frac{\bar{\nu}_1}{A_1}=\frac32\bar{\nu}_1.
$$

From the (\ref{GSexact}), the equation for the mean values follows $(k=0)$:
\begin{equation}
\label{Eaver}
\frac32\bar{\nu}_1^2-\frac38\bar{\nu}_2^2+\frac12\bar{\varrho}^2=\bar{E},
\end{equation}
because $\overline{\nu_1^2}\approx\bar{\nu}_1^2$ an so on. Comparison 
of (\ref{alpha,beta-invar}) and (\ref{Eaver}) gives
$$
\bar{\varrho}=C_s.
$$

These results reveal the following possible cosmological scenarios for the 
Bianchi-IX universe at the semiclassical stage of its evolution according to a 
relation between the parameters $\bar E$ and $C_s$ of the two condensates.

1. Empty space ($C_s=0$, $\bar E_k=0$); 
$\alpha=\frac{\beta}2$.

In the limit $t=\pm\infty$
$$
a^2=b^2=\frac{\beta}8;\;c^2=0, 
$$
i.e. the metric (\ref{ds}) asymptotically takes the form
$$
ds^2=(\beta c^2)dt^2-\frac{\beta}8(d\vartheta^2+\sin^2\vartheta 
d\varphi^2),
$$
$$\vartheta=x^1,\;\;\varphi=x^2.$$

When reaching singularity in one of the dimensions, two others form a 
stationary space of constant curvature. Here one deals with a regime of 
dynamical compactification, a space with simple topology being 
compactified.

2. Space is filled with the medium having positive energy density:
$\bar E_k<\frac12C_s^2;\;\alpha>\frac12\beta$.

For $\alpha=\beta,\;t_0=0$ the model is isotropic.

For $\alpha=\beta,\;t_0\neq0$ the model is unisotropic, but the singularity 
has an isotropic nature.

For $\frac12\beta<\alpha<\beta$ in the pre-singular state $a^2=b^2\gg c^2$, 
i.e. (2 + 1)-dimensional space-time arises, where the 2-space has constant 
curvature.

For $\alpha>\beta$ in the pre-singular state $c^2\gg a^2=b^2$, however, the 
model is not reduced to the space of less dimensions.

In all the cases for $\bar E_k<\frac12C_s^2$ space at singularity is 
contracted to a point.
	
3. Space is filled with the medium having negative energy density: 
$\bar E_k>\frac12C_s^2;\;\alpha<\frac12\beta$.

At $t=\pm\infty$ the third space dimension is compactified ($c^2\to0$), and 
the remaining two-dimensional space of constant curvature is infinitely 
expanding. In the special case $\alpha=\beta/4$ the scale factor $a=b$ 
is increasing exponentially in proper time.

So, the GVC, affecting coupling between the constants $\alpha$ and $\beta$ 
through the controlling parameter $\bar E_k$, determines a cosmological 
scenario 
which may contain the following phenomena:
\begin{itemlist}
\item cosmological expansion and contraction of space;
\item cosmological singularity;
\item compactification of space dimensions;
\item asymptotically stationary space of less dimensions; 
\item inflation of the Universe.
\end{itemlist}
One can see that even such a simplified model reveals a number of effects 
probable from the standpoint of the modern cosmological ideas. Introduction 
of the GVC to the theory enlarges the number of possible cosmological 
scenarios, a concrete value of the parameter $E_k$  being formed at the 
quantum stage of the Universe existence.

Evidently, every classical cosmological evolution scenario must be in 
correspondence with some configuration of the wave packet (\ref{GSexact}) 
-- (\ref{exactE-pack}). But not all the solutions to the SE describe classical 
universes, i.e. form sufficiently narrow packets to satisfy the conditions 
(\ref{mu,nu,omega}); in addition, even those wave packets, for which a 
transition to the semiclassical regime is possible, may prove to be unstable. 
Therefore in our approach the known problem of initial conditions for 
classical evolution is formulated as the problem of choice of the Universe 
quantum state. A quantum state in the Bianchi-IX model is determined by a 
concrete kind of the function
 $$
\tilde{C}(\nu_1,\nu_2,\varrho,E_k)=c_2(E_k)c_1(\nu_1,\nu_2,\varrho),
$$
describing a wave packet structure.

We do not know the way the choice of the quantum state is being made by. 
Perhaps, it is realized according to statistical laws in the process of the 
Universe creation from ``Nothing". In the next paper we intend to discuss 
the hypothesis according to which the act of the Universe creation is 
understood as, occurring out of time, quantum transition from the special 
singular state to one of the Universe physical states which the wave packets 
(\ref{GSexact}) -- (\ref{exactE-pack}) correspond to.

\section{Conception of Time}
\label{time}
\noindent
In the previous section the transition to the semiclassical regime was 
considered in the GVC time. As a matter of fact, there exist four different 
concepts of time in the ordinary quantum theory:
\begin{itemlist}
\item the {\it Heisenberg} time -- the time parameter in Heisenberg 
representation; it is introduced equally with space coordinates by the device 
preparing an initial state of an object ({\it preparator});
\item the {\it Schr\"odinger} time introduced by the registrating device({\it 
registrator});
\item the time in the {\it interaction} representation defined by preparators 
and registrators in the asymptotical  regions of space-time;
\item the world time defined on the object itself in the semiclassical regime 
({\it semiclassical proper} time).
\end{itemlist}

In QGD the only clock carrier is a GVC, and the question about availability 
of the Heisenberg time is closely related to the question about the Universe 
creation. As to the semiclassical proper time (SPT), its existence is not predetermined by 
the theory for, as it was mentioned above, a semiclassical regime is not 
predetermined itself. But, having such a regime actually, let us see how to 
deal with the SPT in the present theory.

From the very beginning we should emphasize that the SPT has nothing to do 
with the GVC as the OM carrier. By transition to the semiclassical regime 
the integrity effects are weakening; as a result, the fact of the GVC existence 
shows only through the parameters responsible, in our case, for the wave 
packet satisfactory width. So, the SPT does not depend on the GVC time, and 
the question, whether we can co-ordinate {\it our} semiclassical clocks with 
the GVC ones, seems to be of considerable physical significance in view of 
obtaining information about the quantum stage of the Universe evolution.

A semiclassical wave packet may also be presented in the form
\begin{eqnarray*}
\Psi_k(z_1,z_2,\phi,t)&=&\int\!\!\!\!\int\limits_{-\infty}^{\infty}\!\!\!\!
\int dE_kd\varrho d\nu_1d\nu_2\tilde{C}(\nu_1,\nu_2,\varrho,E_k)\nonumber\\
&\times&\exp\left[-iE_k(t-t_0)+i\sigma_1(z_1)+i\sigma_2(z_2)\right]\nonumber\\
&\times&\psi_{\varrho}(\phi)\delta\left(\frac32\nu_1^2-\frac38\nu_2^2+
\frac12\varrho^2-E_k\right).
\end{eqnarray*}
Here the sum $\sigma_1(z_1)+\sigma_2(z_2)$ is the part of the classical 
action $S(z_1,z_2,\phi,t)$, determining its dependence on $z_1$ and $z_2$ 
(the scalar field is treated as essentially quantum). The functions $\sigma(z)$ 
satisfy the equations
\begin{equation}
\label{sigma}
\frac{\partial\sigma}{\partial z}=\sqrt{\frac{\bar{\nu}^2}4-
\frac{A^2}4\exp(z)}.
\end{equation}

Knowing the dependence of the classical action on the variables $z$, one can 
reconstruct the evolution law (\ref{solutions}) with the help of the standard 
procedure. But the two mentioned methods of going over to the classical 
limit are applicable owing to the explicit dependence of the GS on time. And 
this, in turn, is caused by the available, in the theory, indication of the 
concrete choice of an RS which the time $t$ is measured in.

In the classical limit a classical subsystem of the physical object itself can 
be considered as an RS. Such a subsystem cannot fill the whole space; it is 
admissible that it would occupy a limited region of space. So, we will refer 
to such an RS as to a local one. The appearance of the time $\tau$, introduced 
as 
a parameter along a classical path, is associated with that RS.

A derivative with respect to path length can be defined by the following way: 
$$\frac d{dt}=u(\tau)\nabla S\nabla,$$
where $u(\tau)$ is an arbitrary function, $\nabla S$ is a tangent vector to 
the path;
$$
\nabla=\left(2\sqrt{3}\,\frac{\partial}{\partial z_1},\;\;i\sqrt{3}\, 
\frac{\partial}{\partial z_2}
\right).
$$
On the other hand,
$$
\frac d{d\tau}=\frac{d z_1}{d\tau}\frac{\partial}{\partial z_1}+
\frac{d z_2}{d\tau}\frac{\partial}{\partial z_2},
$$
whence
$$
\frac{d z_1}{d\tau}=12u(\tau)\frac{d\sigma_1}{d z_1};\;\;
\frac{d z_2}{d\tau}=-3u(\tau)\frac{d\sigma_2}{d z_2}.
$$
From (\ref{sigma}) one obtaines
$$
\frac{d z_1}{d\tau}=4u(\tau)\sqrt{\beta^2-\exp(z_1)};\;\;
\frac{d z_2}{d\tau}=-4u(\tau)\sqrt{\alpha^2-\exp(z_2)},
$$
and, in the result of integration,
\begin{equation}
\label{tau-solut}
\exp\left(\displaystyle\frac{z_1}2\right)=\beta\cosh^{-1} 
\left(2\beta\left[\tilde{u}
(\tau)-\tau'_0\right]\right);\;\;\exp\left(\displaystyle\frac{z_2}2\right)=\alpha
\cosh^{-1}\left(2\alpha\left[\tilde{u}(\tau)-\tau_0\right]\right);
\end{equation}
$$
\tilde{u}(\tau)=\int u(\tau)\,d\tau;\;\;\tau_0,\tau'_0={\,\rm const}.
$$

The time $\tau$ of a local observer emerges irrespectively of the time $t$ 
existence, but both the times may correlate between them. To bring the 
expressions (\ref{solutions}) and (\ref{tau-solut}) in correspondence, it is 
sufficient to put $u(\tau)=1$. This act assumes the following operations:
\begin{romanlist}
\item determining the rate of the Universe evolution by means of 
cosmological observations;
\item fixing, on this data base, the GVC parameters and the state equation of 
it;
\item reconstruction of the gauge from the state equation and so fixing the 
GVC time scale.
\end{romanlist}

In the ordinary quantum mechanics the time variable involving in SE may fail
to coincide with that appearing in Heisenberg operator equations, as well as 
with the world time in which semiclassical system dynamics can be described. 
In the quantum mechanics the hypothesis about equivalence of the mentioned 
time variables is used, though this is nowhere specified. In the considered 
example we have manifested that the time variables used for describing a 
quantum system evolution are different in general. In QGD times associated 
with different observers can be brought to agreement with each other by
choice of a gauge condition.

\nonumsection{References}

\end{document}